\documentclass[conference]{IEEEtran}
\def\BibTeX{{\rm B\kern-.05em{\sc i\kern-.025em b}\kern-.08em
    T\kern-.1667em\lower.7ex\hbox{E}\kern-.125emX}}
\usepackage{cite}
\usepackage{amsmath,amssymb,amsfonts}
\usepackage{graphicx}
\usepackage{textcomp}
\usepackage{algpseudocode}
\usepackage{booktabs}
\usepackage[caption=false,font=footnotesize]{subfig}
\usepackage{multirow}
\usepackage{pifont}
\usepackage{xcolor}
\usepackage{pgfplots}
\pgfplotsset{compat=1.18}
\title{PSP: Low-Overhead Packet-Level Load Balancing for Stale-State and Bandwidth-Asymmetric Networks}

\author{%
\IEEEauthorblockN{Jiaqi Liu\IEEEauthorrefmark{1},
Chunyang Zhang\IEEEauthorrefmark{2},
Heng Pan\IEEEauthorrefmark{1}, and
Yanbiao Li\IEEEauthorrefmark{1}}
\IEEEauthorblockA{\IEEEauthorrefmark{1}Computer Network Information Center, Chinese Academy of Sciences, China}
\IEEEauthorblockA{\IEEEauthorrefmark{2}Huawei Technologies Company Ltd., Shenzhen, China}
}
\begin{document}
\maketitle

\begin{abstract}
With the rapid growth of large language model training and generative artificial intelligence services, data center networks face severe micro-burst traffic and high concurrency. Traditional hash-based flow-level load balancing cannot sense link states, leading to hash collisions, hotspot congestion, and tail latency in multipath Clos networks. Existing packet-level schemes are constrained by stale state information, high hardware complexity, and poor adaptation to heterogeneous links.
To address these issues, this paper proposes probabilistic state-proportional (PSP) dispatching, a packet-level load balancing algorithm. Using a Band-based discrete state representation, PSP replaces global sorting with local probability mapping, reducing hardware complexity while suppressing herding and oscillations caused by stale states.
Experiments on a cycle-accurate simulator show that PSP is robust across port scales, bandwidth-limited paths, and fixed-flow interference. It outperforms join-the-shortest-queue (JSQ) scheduling and Random in loss rate, 99th-percentile buffer occupancy, and scalability, while remaining competitive with Top-$k$ at lower hardware cost. PSP provides an effective balance among performance, stability, and overhead for artificial intelligence data centers.
\end{abstract}

\begin{IEEEkeywords}
Packet-level load balancing, data center networks, stale state information, bandwidth asymmetry, low hardware overhead
\end{IEEEkeywords}

\section{Introduction}
With the explosive growth of large language models (LLMs) and generative AI, data centers are facing unprecedented networking challenges. Distributed AI training workloads rely heavily on collective communication primitives such as All-Reduce and All-to-All\cite{Aeon,Horovod}, and the resulting traffic is characterized by strong burstiness and high concurrency\cite{NDP2017,Homa2018}. In multipath Clos topologies, traditional hash-based flow-level load balancing schemes such as ECMP cannot sense link states. As a result, they are prone to hash collisions when handling elephant flows, which in turn leads to severe hotspots and long-tail latency and ultimately wastes expensive GPU cluster compute resources\cite{ECMP,CONGA,pFabric2013,PIAS2015}.

To extract as much throughput as possible from the network, industry practice is shifting load balancing from coarse-grained flow-level scheduling to finer-grained flowlet-level and packet-level scheduling\cite{HULA2016,LetFlow2017,FlowBender2014,FlowDyn2019,PLB2022}. Although advances in end-host reordering tolerance and switch-side reordering buffers have largely removed the protocol barrier to packet spraying\cite{Presto,DRB}, realizing efficient, stable, and deployable packet-level load balancing in real AI data centers still faces three key challenges.

First, \textbf{adaptation to heterogeneous network environments} remains inadequate. In real deployments, link failures, incremental upgrades, the coexistence of different generations of NICs and switching devices, and long-lived background flows occupying part of the network are all common\cite{Hermes,SAPS,SGLB}. These factors lead to persistent asymmetry in available bandwidth across different paths. Under such conditions, a load balancing algorithm must distinguish not only which path is currently less congested, but also which path is intrinsically weaker in available capacity; otherwise, the scheduler may continue injecting traffic into constrained paths, causing buffer buildup, packet loss, retransmissions, and ultimately slower training jobs.

Second, \textbf{the timeliness of remote state information is difficult to guarantee}. Existing distributed load-sensing and network-telemetry mechanisms typically rely on probes, feedback, or telemetry to propagate remote path states across nodes. For example, HULA disseminates path-utilization information through periodic probes\cite{HULA2016}, while INT supports carrying state metadata such as hop latency, queue occupancy, and buffer occupancy in packets\cite{INT}. However, all such remote states must undergo multi-hop propagation, processing, and updating before reaching the source node, which means they inevitably arrive with non-zero delay. For AI training traffic that is highly sensitive to micro-bursts, even such modest delay is sufficient to trigger concentrated following, local congestion, packet loss, and retransmissions, thereby significantly amplifying collective communication completion time.

Finally, \textbf{the scheduling algorithm must have low hardware overhead}. Packet-level load balancing must operate at line rate in the switch data plane, whereas software-based schemes are constrained by the control-plane feedback loop formed by remote-state export and candidate-port reinstallation, which typically introduces tens to hundreds of microseconds of additional delay and makes the candidate set stale before it even takes effect\cite{HULA2016,P4RuntimeUpdate}. Therefore, port-quality quantization and path selection must be implemented directly in hardware. At the same time, modern high-radix switches have extremely tight silicon-area, routing, and timing budgets, so reducing the hardware complexity of load-balancing logic is essential to ensuring scalability and deployability\cite{TopSort2022,TopKSorter2022,SGLB}.

To address these three challenges, we propose PSP (Probabilistic State-Proportional), a highly robust packet-level load balancing algorithm. The design philosophy of PSP is to replace global sorting with local perception and to replace hard truncation with probabilistic smoothing. Building on Band-based discrete state representation, PSP introduces a state-proportional probabilistic dispatch mechanism that reduces the complex global sorting problem to a set of parallel local mapping tasks.

The main contributions of this work are summarized below:
\begin{itemize}
\item \textbf{A probabilistic PSP dispatch architecture.} Building on Band-based discrete state representation, we design a hardware-efficient probabilistic dispatch logic to mitigate traffic oscillations under stale state information.
\item \textbf{A practical parameter configuration model.} We provide a heuristic analysis of linear probability-threshold mapping and further derive an engineering-oriented model for setting the maximum threshold $Th_{\max}$, reducing complex tuning to direct alignment with end-host hardware capabilities.
\item \textbf{Large-scale cycle-accurate evaluation.} On a 51.2T in-house Clos simulation platform, PSP consistently outperforms JSQ and Random in loss rate, buffer utilization, and scalability, while slightly outperforming Top-$k$ at lower hardware cost overall.
\end{itemize}

\section{Motivation and Problem Definition}

\noindent Packet-level scheduling has regained attention in AI training and high-speed storage networks because it can exploit Clos multipath bandwidth more effectively than flow-level or flowlet-level schemes and thus mitigate local hotspots caused by hash collisions\cite{DRB,Presto,PLB2022}. However, this finer scheduling granularity also forces the algorithm to confront state freshness, hardware implementation complexity, and path-capability differences more directly. For the mixed packet-flow data center setting considered in this paper, existing representative schemes still fail to satisfy all three requirements simultaneously.

\subsection{The Hardware Bottleneck of Top-$k$}
The core idea of Top-$k$ is to first identify a set of currently best output ports from a large candidate set and then choose one port from this reduced set. This strategy leverages state information while providing some robustness, but its cost is that large-scale comparison and selection must be completed within the very tight latency budget of the switch ASIC pipeline. For high-radix switches, this process typically relies on hardware sorting or Top-$k$ selection structures, whose critical-path depth, on-chip resource consumption, and routing overhead grow rapidly with the number of ports\cite{TopSort2022,TopKSorter2022}.

Therefore, the main limitation of Top-$k$ does not lie in the scheduling idea itself, but in its heavy demand for comparators, multiplexers, registers, and cross-stage wiring\cite{TopSort2022,TopKSorter2022}. As port counts continue to increase, sorting depth, chip area, and routing congestion together become a major implementation bottleneck, making Top-$k$ difficult to deploy as a sufficiently lightweight general-purpose packet-level scheduling mechanism.

\subsection{The Sensitivity of JSQ to Stale State}
JSQ and related state-aware schemes typically achieve good load balancing when information is fresh, because they always direct traffic toward shorter queues\cite{JSQMany2018}. In large high-speed networks, however, state collection, upload, and dissemination inevitably incur delay, so the scheduler often sees queue states from several microseconds or even tens of microseconds earlier. Under such conditions, what would otherwise be an effective greedy decision turns into a concentrated chase after stale information, thereby inducing classic herding and divergent oscillations\cite{JSEQDelay2021}.

\begin{figure}[!htbp]
 \centering
 \includegraphics[width=0.92\linewidth]{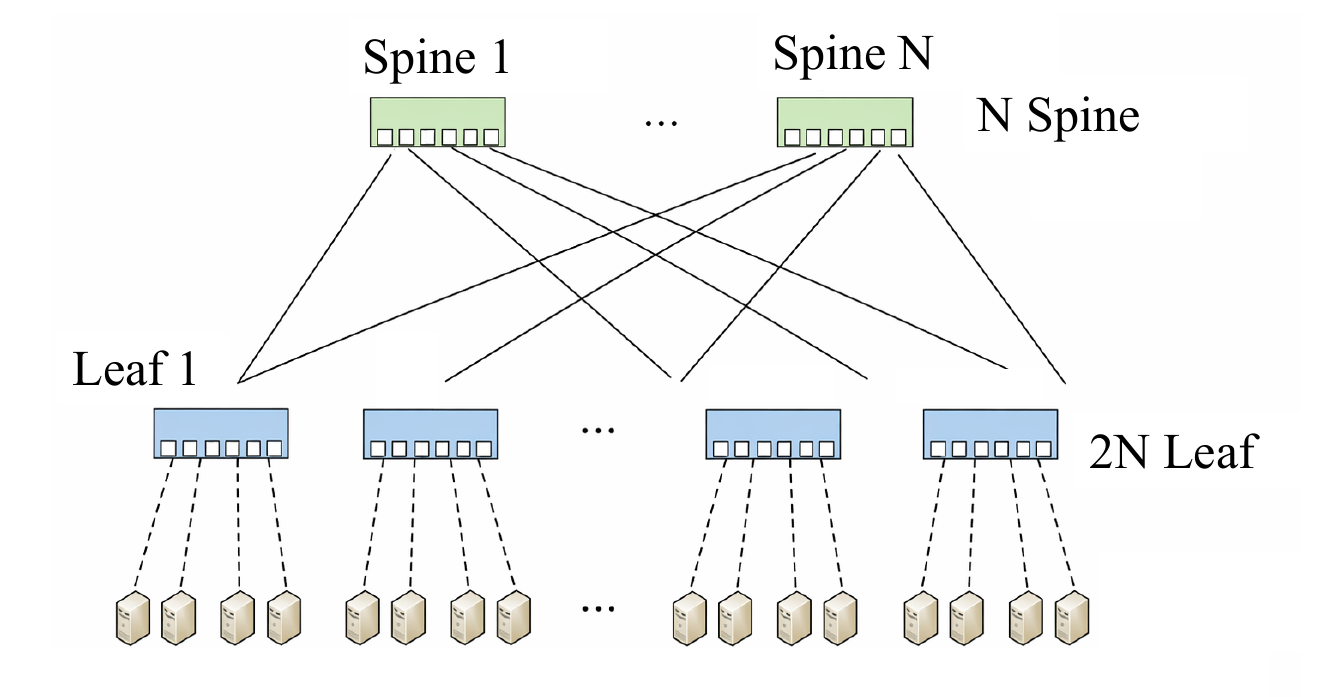}
 \caption{Clos topology used in the simulation platform (two-layer spine-leaf fat-tree architecture).}
 \label{fig:topology}
\end{figure}

As shown in Fig.~\ref{fig:topology}, in a two-layer spine-leaf Clos network, a source ToR making a local path-selection decision cannot synchronously observe the real-time congestion state of remote spine switches or downstream links, and must instead rely on periodic control-plane reports. Thus, the fundamental limitation of JSQ is not its inability to use state, but its excessive sensitivity to state freshness. Once information becomes stale, its high-gain feedback can easily amplify local fluctuations into persistent oscillations\cite{JSEQDelay2021}.

\subsection{Random Is Blind to Path Capability}
Beyond state delay, practical data center networks often operate under mixed packet-flow traffic, link failures, incremental upgrades, and mixed old/new devices\cite{Hermes,SGLB}. These factors cause some paths to remain persistently constrained by fixed flows or lower-rate links, resulting in pronounced asymmetry in available bandwidth. In such an environment, a scheduler must distinguish not only whether a path is currently congested, but also whether the path itself is fundamentally weaker.

Random has the advantage of being simple and insensitive to state freshness, but it also cannot recognize path capability differences. When some paths are already occupied by fixed flows or their effective rates are reduced by heterogeneous links, random dispatching still injects traffic into them in an approximately uniform way. This aggravates local queuing, increases reordering pressure, and reduces multipath bandwidth utilization\cite{DRB,REPS,QDAPS2021,HTPC2021}.

\subsection{Design Goals}
The analysis above shows that existing schemes suffer from clear weaknesses along different dimensions: Top-$k$ is constrained by hardware cost, JSQ is sensitive to stale state, and Random cannot exploit path capability differences. To summarize how representative schemes relate to the three key requirements considered in this paper, Table~\ref{tab:motivation-compare} compares typical algorithms against these requirements.

\begin{table*}[!t]
\caption{Relationship between representative algorithms and the three major challenges addressed in this paper.}
\label{tab:motivation-compare}
\centering
\begin{tabular}{lccc}
\toprule
\textbf{Algorithm} & \textbf{Robust to Stale State} & \textbf{Hardware-Friendly} & \textbf{Adapts to Bandwidth Asymmetry} \\
\midrule
Random & \ding{51} & \ding{51} & --- \\
JSQ & --- & \ding{51} & \ding{51} \\
Top-$k$ & \ding{51} & --- & \ding{51} \\
PSP & \ding{51} & \ding{51} & \ding{51} \\
\bottomrule
\end{tabular}
\end{table*}

\section{PSP Design}

\subsection{Overall Objective}
The primary goal of this work is to design a load balancing algorithm that remains highly robust even under extreme network-state delays. Specifically, the algorithm must satisfy the following requirements:
\begin{enumerate}
\item \textbf{Delay tolerance and generalization.} It should maintain strong load balancing performance across both highly stale control-plane conditions (state update delay $>20\,\mu\text{s}$) and near-real-time settings (state update delay $<1\,\mu\text{s}$), while approaching the ideal-performance baseline of JSQ.
\item \textbf{Low hardware resource overhead.} Under the stringent forwarding-rate demands of switch chips, the algorithm's computation and storage overhead must be tightly controlled to match the hardware constraints of modern commercial data center switches.
\item \textbf{Adaptation to complex scenarios.} Under bandwidth-constrained paths, asymmetric topologies, and 100\% offered load, the algorithm must continuously avoid constrained ports and sustain balanced traffic distribution.
\end{enumerate}

\subsection{Hardware Architecture and Implementation of PSP}
Unlike Top-$k$, which depends on global sorting, PSP does not pursue exact comparison across all port states in hardware. Instead, it first discretizes continuous queue depth into a finite set of load states and then completes scheduling through local mapping followed by global weighted selection. In this way, the tightly coupled sorting problem is transformed into a combination of local per-port processing and logarithmic-depth global aggregation, which is better suited to parallel pipelined implementation on high-radix switches.

\begin{figure*}[t]
  \centering
  \includegraphics[width=0.95\textwidth]{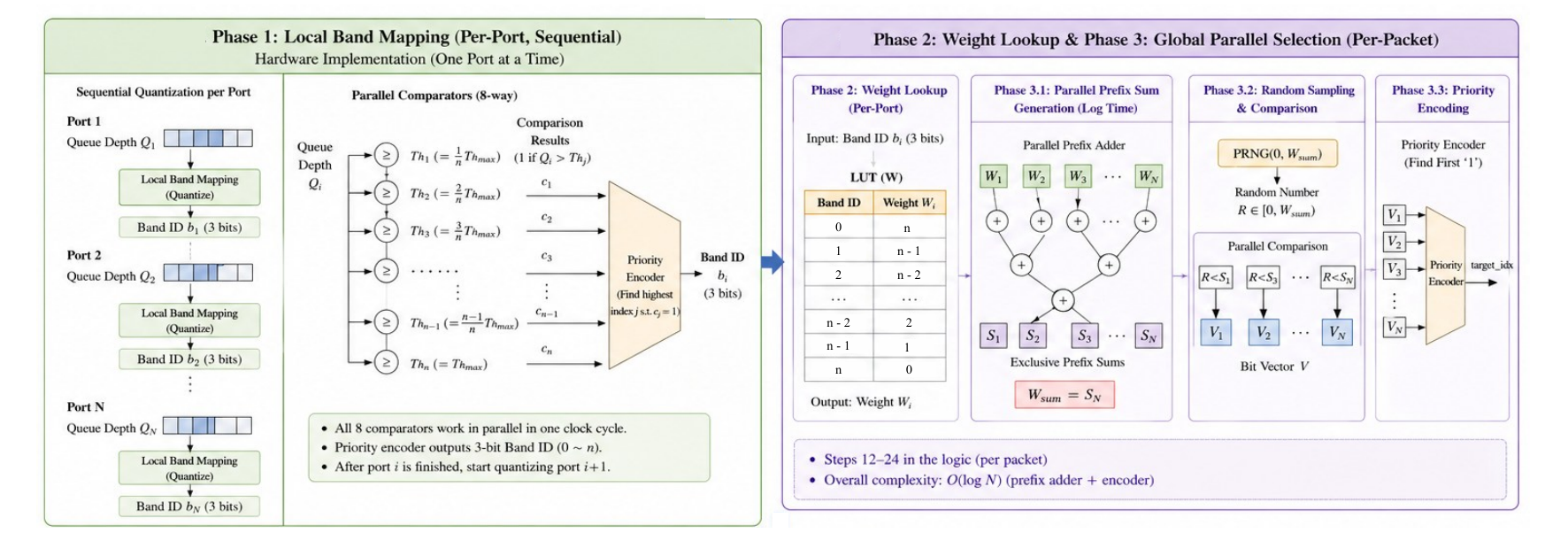}
  \caption{Hardware architecture of PSP: local interval mapping, probabilistic weight lookup, and global weighted selection.}
  \label{fig:psp}
\end{figure*}

As shown in Fig.~\ref{fig:psp} and the pseudocode in Fig.~\ref{fig:psp-logic}, the hardware architecture of PSP consists of three sequential stages. The first stage is Local Band Mapping: each port compares its own queue depth with the threshold vector and quantizes the continuous state into a discrete Band ID. This process avoids inter-port state exchange while naturally filtering short-timescale micro-burst perturbations through discretization.

The second stage is Weight Assignment. After obtaining the Band ID, each port maps the discrete state to its corresponding probabilistic weight $W_i$ through a local lookup table. This mapping allows PSP to change scheduling preference by modifying the weight configuration without altering the overall hardware structure, thereby combining implementation simplicity with parameter tunability.

The third stage is Global Weighted Selection. After all port weights are aggregated, the system first uses a parallel prefix sum to generate weight boundaries, then performs pseudo-random sampling and parallel comparison to identify the matched interval, and finally outputs the target port through a priority encoder. Since this process is built on regular adders and comparators, its critical path remains at $O(\log N)$, avoiding the high coupling and routing overhead of global sorting networks.

Overall, PSP converts load balancing decisions into hardware operations that can be decomposed and parallelized through a three-stage pipeline of local quantization, weight mapping, and weighted selection. This significantly reduces implementation complexity while retaining state awareness. The hardware benefits are further quantified in the next subsection.

\subsection{Hardware Resource Overhead Analysis}

To evaluate the hardware efficiency of PSP, we compare it against a traditional Top-$k$ selection design represented by a bitonic sorting network on a 128-port high-radix switch.

\subsubsection{Local Processing Stage}
\textbf{Parallel quantization vs.\ sorting elements.} In a Top-$k$ design, the core consists of a large number of compare-and-swap (CAS) units. For 128 ports, a bitonic sorting network requires $\frac{N \log_2 N (\log_2 N + 1)}{4}$ CAS units, i.e., 1,792 units. Each unit includes a wide comparator, a multiplexer (MUX), and complex swap logic. PSP instead adopts Local Band Mapping, where each port only needs a static comparator array (quantizer) and a small lookup table (LUT). Since the logic of different ports is completely decoupled and no data exchange is required, the basic gate count is reduced by roughly 90\% relative to Top-$k$.

\subsubsection{Global Aggregation Stage}
\textbf{Adder tree vs.\ sorting network.} In Top-$k$, sorting is itself the decision process, so the physical routing of the sorting network exhibits a nonlinear all-to-all structure, producing huge routing-area overhead and logical dead space. PSP instead aggregates weights with a parallel prefix adder tree. For $N$ ports, only $N-1$ adders are required; when $N=128$, this is just 127 adders. The adder-tree structure is regular and unidirectional, greatly improving placement and routing density.

\subsubsection{Auxiliary Logic and Register Overhead}
\textbf{Register redundancy.} Because the combinational logic depth of Top-$k$ is large, thousands of pipeline registers must be inserted to cut timing paths and achieve timing closure at GHz frequency, resulting in substantial extra area overhead.

\textbf{Random selection logic.} Although PSP introduces a pseudo-random number generator (PRNG), such as an LFSR, and a weighted selector, the physical scale of this logic is small and requires only tens of flip-flops.

\begin{table*}[!t]
\caption{Estimated hardware resource comparison between PSP and Top-$k$ for a 128-port switch (coarse-grained estimate based on standard-gate equivalence).}
\label{tab:hw-resource-compare}
\centering
\footnotesize
\begin{tabular}{p{2.8cm}p{4.4cm}p{4.4cm}p{3.3cm}}
\toprule
\textbf{Stage} & \textbf{Top-$k$} & \textbf{PSP} & \textbf{Relative Estimate} \\
\midrule
Port-state processing & Dense CAS array, complexity about $O(N\log^2 N)$ & Parallel quantizers + local LUTs, complexity about $O(N)$ & PSP area reduced by about $\sim 90\%$ \\
Global decision logic & Sorting-network interconnect with severe global congestion & Binary prefix adder tree + PRNG, regular structure & PSP area reduced by about $\sim 85\%$ \\
Pipeline and register overhead & Multi-stage pipeline registers required for timing closure & Shorter timing paths and lower register overhead & PSP overhead reduced by about $\sim 80\%$ \\
\midrule
Overall estimate & High area and severe routing congestion & Low area and linear scalability & PSP total area is about $\sim 12\%$ of Top-$k$ \\
\bottomrule
\end{tabular}
\end{table*}

\subsubsection{Summary}
By converting load balancing decisions into parallel local mapping tasks, PSP fundamentally breaks the ``sorting wall'' encountered by high-radix switch designs. Its $O(\log n)$ critical path greatly simplifies hardware implementation and also reduces the additional staleness caused by hardware latency itself, laying the foundation for ultra-low-latency scheduling at 400\,Gbps and beyond.

\subsection{Robustness to Stale Information}
The probabilistic state-proportional dispatching mechanism of PSP is designed to mitigate load imbalance caused by stale information. In the baseline comparison of this paper, JSQ also adopts Band-based discretization, but its scheduling rule remains strictly greedy: it always selects a port uniformly at random from the set of ports currently in the lowest-load Band.

This ``random within the best Band'' rule still carries a high oscillation risk under stale state information. Compared with JSQ, which concentrates burst traffic onto a small number of currently optimal ports, PSP significantly reduces the instantaneous injected traffic per port within the stale-information window, thereby alleviating local congestion and subsequent oscillations. A concrete example is given in Appendix~\ref{appendix:herding}.

\subsection{Adaptive Regulation Under Bandwidth Asymmetry}
In addition to mitigating herding under stale state information, PSP also exhibits strong self-healing capability in bandwidth-asymmetric environments. Fundamentally, PSP behaves as a negative-feedback control model: it uses queue buildup caused by instantaneous congestion as the feedback signal. When the actual outgoing capacity of a path falls below the expected level, incoming traffic accumulates in the corresponding port buffer. Through Band mapping, this buildup is converted into a lower scheduling weight, which probabilistically suppresses additional traffic sent onto that path and eventually drives the incoming rate toward the residual-bandwidth equilibrium point. A quantitative analysis of this process is provided in Appendix~\ref{appendix:bandwidth-asymmetry}.

\begin{figure}[!t]
\centering
\begin{minipage}{\columnwidth}
\begin{algorithmic}[1]
\Statex \textbf{I. Configuration Stage (Static / Global)}
\State \textbf{Parameter:} $n$ \Comment{Number of Bands, e.g., 8}
\State \textbf{Parameter:} $Th_{max}$ \Comment{Maximum threshold, e.g., 960\,KB}
\State \textbf{Thresholds:} $\mathcal{T} = \{ \frac{Th_{max}}{n}, \frac{2Th_{max}}{n}, \dots, Th_{max} \}$ \Comment{Linear distribution}
\State \textbf{Weights:} $\mathcal{W} = \{ n, n-1, \dots, 1 \}$ \Comment{Linearly decreasing weights}

\Statex
\Statex \textbf{II. Stages 1 \& 2: Local Band Mapping and Weight Lookup (Parallel Across Ports)}
\Procedure{ParallelLocalMapping}{$\text{QueueDepths}[1 \dots N]$}
    \For{\textbf{each} $port \in \mathcal{P}$ \textbf{in parallel}}
        \State $q \gets \text{QueueDepths}[port]$
        \State $b \gets \text{Quantize}(q, \mathcal{T})$ \Comment{Hardware interval comparator array}
        \State $W[port] \gets \text{LUT}(b)$ \Comment{Assign $W_i$ from $\mathcal{W}$ according to $b$}
    \EndFor
\EndProcedure

\Statex
\Statex \textbf{III. Stage 3: Global Parallel Selection (Per Packet)}
\Function{PSPScheduling}{$packet$}
    \Statex \quad \textit{// Sub-stage 3.1: parallel prefix-sum generation (logarithmic time)}
    \State $\mathcal{S}[1 \dots N] \gets \text{ParallelPrefixAdder}(W[1 \dots N])$
    \State $W_{sum} \gets \mathcal{S}[N]$

    \Statex \quad \textit{// Sub-stage 3.2: random sampling and comparison}
    \State $R \gets \text{PRNG}(0, W_{sum})$ \Comment{Generate a hardware random number}
    \State $V[1 \dots N] \gets \mathbf{0}$ \Comment{Initialize a bit vector}
    \For{$i = 1$ \textbf{to} $N$ \textbf{in parallel}}
        \If{$R < \mathcal{S}[i]$} $V[i] \gets 1$ \Else \ $V[i] \gets 0$
        \EndIf
    \EndFor

    \Statex \quad \textit{// Sub-stage 3.3: priority encoding}
    \State $target\_idx \gets \text{PriorityEncoder}(V)$ \Comment{Find the index of the first ``1''}
    \State \Return $target\_idx$
\EndFunction

\Statex
\Function{Quantize}{$q, \mathcal{T}$} \Comment{Hardware interval-classification primitive}
    \State \Return $\sum_{i=1}^{n-1} (q > T_i)$ \Comment{Implemented with a comparator array}
\EndFunction
\end{algorithmic}
\end{minipage}
\caption{Pseudocode of the PSP scheduling logic.}
\label{fig:psp-logic}
\end{figure}

\subsection{Parameter Configuration}
this subsection presents practical parameter configuration guidelines. For PSP, the core parameters include the number of Bands $N$, the Band-partition thresholds, and the probability weights assigned to each Band. If exhaustive search or grid search is used, the joint search space formed by these three groups of parameters quickly becomes impractical. Therefore, we first identify a suitable threshold-probability mapping form through heuristic analysis, then fix the maximum threshold $Th_{\max}$ according to engineering constraints, and finally study the impact of the Band count on performance. The adopted parameter configuration is as follows:

\textbf{Thresholds:} $Th/n,\, 2Th/n,\, 3Th/n,\, \dots,\, n \cdot Th/n$

\textbf{Probabilities:} $n/\mathit{Sum},\, (n-1)/\mathit{Sum},\, \dots,\, 1/\mathit{Sum}$, where $\mathit{Sum} = n(n+1)/2$

The engineering model for the maximum threshold is
\[
Th_{max} = L_{buffer}
\]
Detailed heuristic analysis is given in Appendix~\ref{appendix:config-analysis}.

\section{Evaluation}

\subsection{Experimental Rules and Workloads}
To comprehensively validate the effectiveness and robustness of PSP, we conduct experiments on an in-house packet-level cycle-accurate simulator; the motivation for using this platform is discussed in Appendix~\ref{appendix:platform}. All experiments follow the same rules: the packet scheduling granularity is 4\,KB, the state update period is configurable, and the default setting combines $20\,\mu\text{s}$ stale state information, $1\,\mu\text{s}$ near-real-time information, and 100\% offered load. The workload suite covers two core scenario classes. The first is stale-state testing on deployed Clos networks, which is used to characterize the concentrated scheduling and oscillatory behavior of JSQ under delayed state information. The second is available-bandwidth asymmetry testing under stale-state networks, which reveals the limitation of Random in not sensing link-state differences. The chosen workload style and packet-level metrics are also consistent with recent datacenter transport studies spanning deadline- and tail-sensitive traffic, packetized congestion control, and RDMA-oriented deployments\cite{pFabric2013,PIAS2015,pHost2015,NDP2017,Homa2018,Swift2020,DCQCN2015,IRN2018,ExpressPass2016,Dart2018}. Because the packet reordering risk studied in this paper mainly manifests as path-level queuing-delay differences, and recent packet-spraying studies likewise identify asymmetry-induced delay gaps and queue buildup as the dominant sources of reordering, we use 99th-percentile buffer occupancy as a proxy for reordering pressure\cite{Presto,QDAPS2021,HTPC2021}, together with average loss rate to characterize stability and availability.

\subsection{Simulation Platform}

\subsubsection{Simulator Architecture}
We use an in-house high-precision packet-level cycle-accurate simulator as the experimental platform. The simulator uses packets as the minimum scheduling unit and accurately models switch-port enqueue/dequeue behavior, link propagation delay, and control-plane state update delay. Its main features include: (1) packet-level event-driven scheduling with nanosecond precision; (2) configurable state refresh periods ranging from near-real-time ($<1\,\mu\text{s}$) to highly stale ($>20\,\mu\text{s}$) conditions; (3) programmable load balancing strategy interfaces that support hot switching among JSQ, Random, Top-$k$, PSP, and other algorithms; and (4) a network-stack model that includes both link-layer queuing and physical-layer bandwidth constraints.

\subsubsection{Network Topology and Hardware Parameters}
The network experiments use the two-layer Clos (fat-tree) topology shown in Fig.~\ref{fig:topology}. The detailed configuration is summarized in Table~\ref{tab:sim-config}.

This topology choice is intended to match a common deployment unit rather than to introduce an artificial simplification. Prior datacenter load-balancing work has explicitly argued that two-tier leaf-spine fabrics cover the needs of most enterprise datacenter deployments and therefore form the primary engineering target for fine-grained in-fabric scheduling\cite{CONGA}. Industry deployment guides likewise treat Layer-2 leaf-spine as a standard production design and note that, when the scale of a Layer-3 underlay is unnecessary, an L2 leaf-spine subnet is often sufficient for real deployments\cite{AristaL2LS}. Therefore, evaluating PSP on a two-layer Clos fabric directly targets the practical setting in which packet-level load balancing is most often exercised within a pod or subnet, even if larger clusters are further interconnected through higher-layer routing domains.

Each spine switch is equipped with $2N$ downlink ports, and each leaf connects to all spine switches through uplinks to preserve full redundancy. As a result, the total throughput of each spine is $N \times 400$\,Gbps. All ports share a virtual output queue buffer with dynamic watermark control, which is intended to emulate the on-chip buffering capability of real commercial switch ASICs.

\begin{table}[!b]
\caption{Core configuration parameters of the simulation platform.}
\label{tab:sim-config}
\centering
\footnotesize
\begin{tabular}{lll}
\toprule
\textbf{Category} & \textbf{Parameter} & \textbf{Value} \\
\midrule
\multirow{2}{*}{Topology scale}
& Number of spine switches & $N$ \\
& Number of leaf switches & $2N$ \\
\midrule
\multirow{4}{*}{Link configuration}
& Spine--leaf link rate & 400 Gbps \\
& Leaf--server link rate & 400 Gbps \\
& Downward ports per spine & $2N$ \\
& Total throughput per spine & $400N$ Gbps \\
\midrule
\multirow{3}{*}{Switch parameters}
& Per-port queue buffer & 1024 KB \\
& Scheduling granularity & Packet-level \\
& State update period & 0.5~$\mu\text{s}$--50~$\mu\text{s}$\\
\bottomrule
\end{tabular}
\end{table}

\subsubsection{Traffic Models and Load Configuration}
To cover diverse traffic characteristics in modern data centers, we adopt three representative traffic generation models.

Unless otherwise specified, the default experimental setup uses the AI-training traffic model, 100\% average network load, and a $20\,\mu\text{s}$ state update delay (the stale-state setting). Each simulation runs for 10\,ms and is repeated five times; the reported results are averaged to eliminate random fluctuations.

\subsection{Baseline Algorithms}
To ensure a fair and comprehensive comparison, we choose the following representative load balancing algorithms as baselines:

\begin{table*}[!t]
\caption{Description of baseline algorithms.}
\label{tab:baselines}
\centering
\footnotesize
\begin{tabular}{lp{10.3cm}p{3.4cm}}
\toprule
\textbf{Algorithm} & \textbf{Mechanism} & \textbf{Key Parameter} \\
\midrule
JSQ & Always selects a forwarding port from the currently best Band (a purely greedy packet-level strategy built on Band discretization) & Global Band visibility \\
Random & Fully stateless uniform random packet dispatching, unable to sense any link state & None \\
Top-$k$ & Selects the $k$ output ports with the shortest queues and forwards uniformly at random among them & $k$ \\
PSP & The probabilistic state-proportional algorithm proposed in this paper: discrete Band mapping + linear proportional probability weights & $N_{\text{band}}\!=\!8$, $W\!=\!\{7,6,5,4,3,2,1,0\}$ \\
\bottomrule
\end{tabular}
\end{table*}

These baselines cover the dominant design patterns discussed in recent data center load-balancing and transport literature, namely greedy queue-aware selection, oblivious spraying, and candidate-pruned state-aware dispatching\cite{HULA2016,LetFlow2017,FlowBender2014,FlowDyn2019,PLB2022,pFabric2013,PIAS2015,pHost2015,NDP2017,Homa2018,Swift2020,DCQCN2015,IRN2018,ExpressPass2016,Dart2018}.

These baselines are chosen for the following reasons. (1) \textbf{JSQ} represents a purely greedy packet-level strategy based on Band discretization. Because it always selects from the best Band, it is prone to ``many-to-few'' congestion when the number of available ports is small, making it suitable for illustrating the fragility of pure greediness. (2) \textbf{Random} represents the upper bound of a fully stateless scheme and is used to measure the performance gain brought by state awareness. (3) \textbf{Top-$k$} represents a strong and widely adopted advanced configuration in the literature.

\subsection{Experiments in Complex Scenarios}
To validate the robustness of PSP to stale information under realistic network conditions,
 we conduct large-scale experiments on the Clos network shown in Fig.~\ref{fig:topology}. 
 This environment naturally carries about $20\,\mu\text{s}$ of state update delay, 
 so scheduling decisions are always made based on stale queue information. 
We test three scales, 128, 256, and 512 GPUs, and compare Top-$k$ ($k=N/2$), PSP, JSQ, and Random in terms of 99th-percentile buffer occupancy, average buffer occupancy, and packet loss.

\subsubsection{Stale-State Test}
\begin{figure*}[!t]
    \centering
    \subfloat[99th-percentile buffer]{%
    \begin{minipage}{0.48\textwidth}
        \centering
        \begin{tikzpicture}
            \begin{axis}[
                ybar,
                bar width=3pt,
                width=\linewidth,
                height=3.2cm,
                ylabel={99th-Percentile Buffer (MB)},
                symbolic x coords={128,256,512},
                xtick=data,
                xlabel={Number of GPUs},
                ymin=0, ymax=16,
                tick label style={font=\footnotesize},
                label style={font=\footnotesize},
                legend style={at={(0.5,1.18)}, anchor=south, legend columns=4, font=\footnotesize},
                ymajorgrids=true,
                grid style=dashed,
                enlarge x limits=0.18,
            ]
                \addplot coordinates {(128,11.32) (256,13.54) (512,14.34)};
                \addplot coordinates {(128,2.31) (256,2.51) (512,2.79)};
                \addplot coordinates {(128,3.43) (256,2.77) (512,1.96)};
                \addplot[fill=blue!60, nodes near coords, point meta=y, nodes near coords style={/pgf/number format/fixed, /pgf/number format/precision=1}, every node near coord/.append style={rotate=0, anchor=south, yshift=2pt, font=\footnotesize}] coordinates {(128,2.41) (256,1.45) (512,1.34)};
                \legend{JSQ, Random, Top-$k$, PSP}
            \end{axis}
        \end{tikzpicture}
    \end{minipage}}\hfill
    \subfloat[Average buffer]{%
    \begin{minipage}{0.48\textwidth}
        \centering
        \begin{tikzpicture}
            \begin{axis}[
                ybar,
                bar width=3pt,
                width=\linewidth,
                height=3.2cm,
                ylabel={Average Buffer (MB)},
                symbolic x coords={128,256,512},
                xtick=data,
                xlabel={Number of GPUs},
                ymin=0, ymax=2.2,
                tick label style={font=\footnotesize},
                label style={font=\footnotesize},
                ymajorgrids=true,
                grid style=dashed,
                enlarge x limits=0.18,
            ]
                \addplot coordinates {(128,1.93) (256,1.8) (512,1.67)};
                \addplot coordinates {(128,0.34) (256,0.29) (512,0.23)};
                \addplot coordinates {(128,0.69) (256,0.26) (512,0.18)};
                \addplot[fill=blue!60, nodes near coords, point meta=y, nodes near coords style={/pgf/number format/fixed, /pgf/number format/precision=1}, every node near coord/.append style={rotate=0, anchor=south, yshift=2pt, font=\footnotesize}] coordinates {(128,0.71) (256,0.28) (512,0.21)};
            \end{axis}
        \end{tikzpicture}
    \end{minipage}}
    \caption{Results of the stale-state test: buffer and loss behavior of different algorithms at different GPU counts in a Clos network that naturally carries $20\,\mu\text{s}$ stale-state delay.}
    \label{fig:stale_network_charts}
\end{figure*}
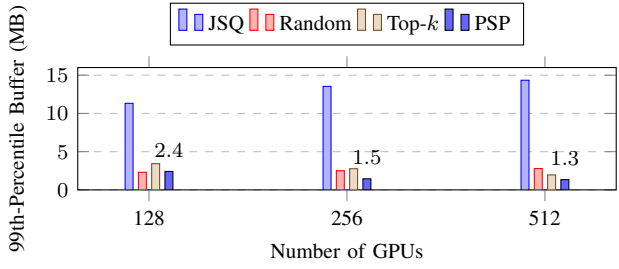
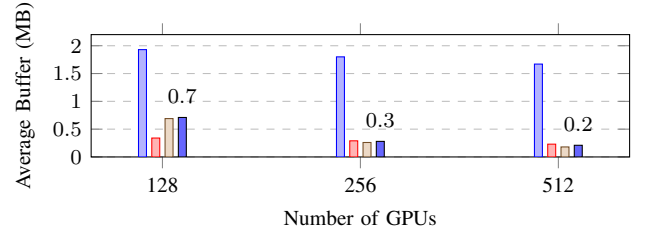

In this scenario, bandwidth remains symmetric: all links operate at 400\,Gbps, and there are no special congestion sources such as link failures or elephant flows. This test is intended to validate the stability and performance advantages of PSP under stale state information. The results are shown in Fig.~\ref{fig:stale_network_charts}.

\begin{table}[!b]
\caption{Packet-loss comparison of all algorithms across the three test settings (\ding{55} indicates packet loss).}
\label{tab:loss-compare}
\centering
\begin{tabular}{lccc}
\toprule
\textbf{Packet Loss} & \textbf{128 GPUs} & \textbf{256 GPUs} & \textbf{512 GPUs} \\
\midrule
Random & \ding{51} & \ding{55} & \ding{55} \\
JSQ & \ding{55} & \ding{55} & \ding{55} \\
Top-$k$ & \ding{51} & \ding{51} & \ding{51} \\
PSP & \ding{51} & \ding{51} & \ding{51} \\
\bottomrule
\end{tabular}
\end{table}

\subsubsection{Available-Bandwidth Asymmetry Tests}
To further evaluate adaptability to available-bandwidth asymmetry, we design two tests: one in which a fixed flow limits local path bandwidth, and another in which physical bandwidth asymmetry is created because part of the network is constrained by lower-speed links.

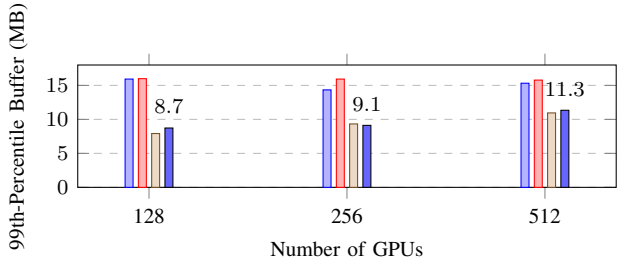
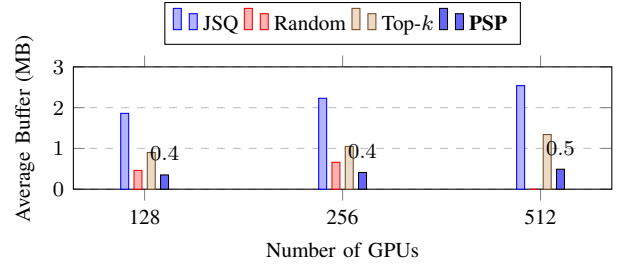
\begin{figure*}[!t]
    \centering
    \subfloat[99th-percentile buffer]{%
    \begin{minipage}{0.48\textwidth}
        \centering
        \begin{tikzpicture}
            \begin{axis}[
                ybar,
                bar width=3pt,
                width=\linewidth,
                height=3.2cm,
                ylabel={99th-Percentile Buffer (MB)},
                symbolic x coords={128,256,512},
                xtick=data,
                xlabel={Number of GPUs},
                ymin=0, ymax=18,
                tick label style={font=\footnotesize},
                label style={font=\footnotesize},
                ymajorgrids=true,
                grid style=dashed,
                enlarge x limits=0.18,
            ]
                \addplot coordinates {(128,15.93) (256,14.34) (512,15.32)};
                \addplot coordinates {(128,15.99) (256,15.93) (512,15.78)};
                \addplot coordinates {(128,7.92) (256,9.33) (512,10.95)};
                \addplot[fill=blue!60, nodes near coords, point meta=y, nodes near coords style={/pgf/number format/fixed, /pgf/number format/precision=1}, every node near coord/.append style={rotate=0, anchor=south, yshift=2pt, font=\footnotesize, text=black}] coordinates {(128,8.73) (256,9.12) (512,11.34)};
            \end{axis}
        \end{tikzpicture}
    \end{minipage}}\hfill
    \subfloat[Average buffer]{%
    \begin{minipage}{0.48\textwidth}
        \centering
        \begin{tikzpicture}
            \begin{axis}[
                ybar,
                bar width=3pt,
                width=\linewidth,
                height=3.2cm,
                ylabel={Average Buffer (MB)},
                symbolic x coords={128,256,512},
                xtick=data,
                xlabel={Number of GPUs},
                ymin=0, ymax=3,
                tick label style={font=\footnotesize},
                label style={font=\footnotesize},
                legend style={at={(0.5,1.18)}, anchor=south, legend columns=4, font=\footnotesize},
                ymajorgrids=true,
                grid style=dashed,
                enlarge x limits=0.18,
            ]
                \addplot coordinates {(128,1.86) (256,2.23) (512,2.54)};
                \addplot coordinates {(128,0.46) (256,0.66) (512,0.0)};
                \addplot coordinates {(128,0.90) (256,1.05) (512,1.34)};
                \addplot[fill=blue!60, nodes near coords, point meta=y, nodes near coords style={/pgf/number format/fixed, /pgf/number format/precision=1}, every node near coord/.append style={rotate=0, anchor=south, yshift=2pt, font=\footnotesize, text=black}] coordinates {(128,0.35) (256,0.41) (512,0.49)};
                \legend{JSQ, Random, Top-$k$, \textbf{PSP}}
            \end{axis}
        \end{tikzpicture}
    \end{minipage}}
    \caption{Results under available-bandwidth asymmetry caused by a fixed flow: the network contains a 200\,Gbps fixed flow along tor15--spine8--tor10.}
    \label{fig:fixed_flow_asymmetry}
\end{figure*}

\paragraph{Fixed-flow scenario}
In this test, the network contains a fixed 200\,Gbps flow along the path tor15--spine8--tor10, while all remaining traffic follows the normal port-competition pattern. This fixed flow persistently occupies local path resources and therefore creates stable asymmetry in available bandwidth. The results are shown in Fig.~\ref{fig:fixed_flow_asymmetry}.

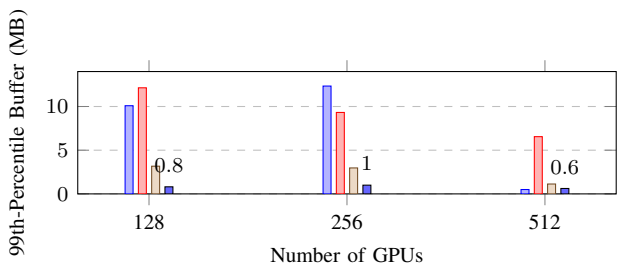
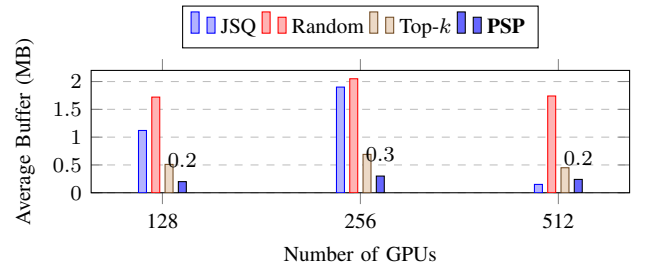
\begin{figure*}[!t]
    \centering
    \subfloat[99th-percentile buffer]{%
    \begin{minipage}{0.48\textwidth}
        \centering
        \begin{tikzpicture}
            \begin{axis}[
                ybar,
                bar width=3pt,
                width=\linewidth,
                height=3.2cm,
                ylabel={99th-Percentile Buffer (MB)},
                symbolic x coords={128,256,512},
                xtick=data,
                xlabel={Number of GPUs},
                ymin=0, ymax=14,
                tick label style={font=\footnotesize},
                label style={font=\footnotesize},
                ymajorgrids=true,
                grid style=dashed,
                enlarge x limits=0.18,
            ]
                \addplot coordinates {(128,10.09) (256,12.34) (512,0.50)};
                \addplot coordinates {(128,12.14) (256,9.32) (512,6.55)};
                \addplot coordinates {(128,3.17) (256,2.97) (512,1.13)};
                \addplot[fill=blue!60, nodes near coords, point meta=y, nodes near coords style={/pgf/number format/fixed, /pgf/number format/precision=1}, every node near coord/.append style={rotate=0, anchor=south, yshift=2pt, font=\footnotesize}] coordinates {(128,0.81) (256,1.00) (512,0.62)};
            \end{axis}
        \end{tikzpicture}
    \end{minipage}}\hfill
    \subfloat[Average buffer]{%
    \begin{minipage}{0.48\textwidth}
        \centering
        \begin{tikzpicture}
            \begin{axis}[
                ybar,
                bar width=3pt,
                width=\linewidth,
                height=3.2cm,
                ylabel={Average Buffer (MB)},
                symbolic x coords={128,256,512},
                xtick=data,
                xlabel={Number of GPUs},
                ymin=0, ymax=2.2,
                tick label style={font=\footnotesize},
                label style={font=\footnotesize},
                legend style={at={(0.5,1.18)}, anchor=south, legend columns=4, font=\footnotesize},
                ymajorgrids=true,
                grid style=dashed,
                enlarge x limits=0.18,
            ]
                \addplot coordinates {(128,1.12) (256,1.90) (512,0.15)};
                \addplot coordinates {(128,1.72) (256,2.05) (512,1.74)};
                \addplot coordinates {(128,0.51) (256,0.69) (512,0.45)};
                \addplot[fill=blue!60, nodes near coords, point meta=y, nodes near coords style={/pgf/number format/fixed, /pgf/number format/precision=1}, every node near coord/.append style={rotate=0, anchor=south, yshift=2pt, font=\footnotesize}] coordinates {(128,0.20) (256,0.30) (512,0.24)};
                \legend{JSQ, Random, Top-$k$, \textbf{PSP}}
            \end{axis}
        \end{tikzpicture}
    \end{minipage}}
    \caption{Results under bandwidth asymmetry caused by heterogeneous links: part of the paths are constrained by lower-speed devices in an incremental deployment setting.}
    \label{fig:heterogeneous_link_asymmetry}
\end{figure*}

\paragraph{Heterogeneous-link scenario}
This scenario models physical bandwidth asymmetry under incremental deployment or mixed generations of devices, where some paths are constrained by lower-speed equipment: specifically, half of the links operate at only half the bandwidth of standard 400\,Gbps links. The results are shown in Fig.~\ref{fig:heterogeneous_link_asymmetry}.

Taken together, these three scenarios show the following. In network environments with stale state information, greedy JSQ performs worst, and its degradation becomes more pronounced as scale increases. Although Random achieves load balancing close to PSP in the pure stale-state scenario, its lack of state awareness causes the most severe imbalance under available-bandwidth asymmetry. Both JSQ and Random suffer noticeable packet loss. By contrast, Top-$k$ ($k=N/2$) maintains strong performance across all three scenarios, indicating that moderately trimming the candidate set helps suppress incorrect following under stale state at larger scales. PSP significantly mitigates stale-state-induced oscillations through probabilistic smoothing. Its buffer occupancy and loss rate remain below those of JSQ and Random and outperform Top-$k$ in most scenarios.

\subsection{Parameter Sensitivity Analysis}
\subsubsection{Impact of the Number of Bands $N$}
The number of Bands $N$ is a core hyperparameter of PSP because it determines the granularity of state discretization. Keeping the maximum threshold fixed at $Th_{\max}=960$\,KB, we test the performance impact of four configurations, $N \in \{4, 8, 16, 32\}$. All configurations use threshold vectors that form arithmetic sequences spanning 64 to 960\,KB. This means that the covered queue-depth range remains identical for all $N$ values, and only the discretization step size changes. This design guarantees a fair comparison across different $N$: any performance difference comes solely from the discretization granularity rather than from changes in threshold coverage.
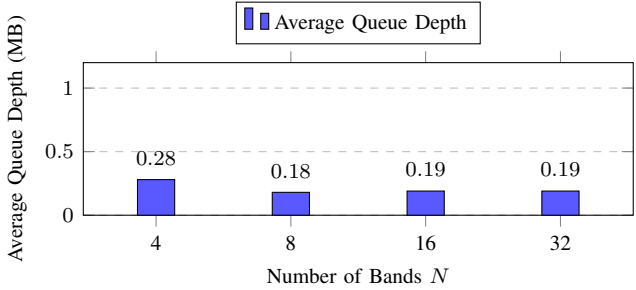
\begin{figure}[!htbp]
\centering
\begin{tikzpicture}
\begin{axis}[
    ybar,
    bar width=14pt,
    ymin=0, ymax=1.2,       
    ylabel={Average Queue Depth (MB)},
    xlabel={Number of Bands $N$},
    symbolic x coords={4,8,16,32},
    xtick=data,
    width=\columnwidth,
    height=3.6cm,
    ymajorgrids=true,
    grid style=dashed,
    nodes near coords,
    point meta=y,
    nodes near coords style={/pgf/number format/fixed, /pgf/number format/precision=2},
    nodes near coords align={vertical},
    enlarge x limits=0.18,
    tick label style={font=\footnotesize},
    label style={font=\footnotesize},
    legend style={at={(0.5,1.12)},anchor=south,legend columns=1,font=\footnotesize},
    every node near coord/.append style={yshift=2pt, font=\footnotesize},
]
\addplot[fill=blue!65] coordinates {(4,0.28) (8,0.18) (16,0.19) (32,0.19)};
\addlegendentry{Average Queue Depth}
\end{axis}
\end{tikzpicture}
\caption{Impact of the number of Bands $N$ on average queue depth. As $N$ increases from 4 to 8, the average queue depth drops significantly; when $N \ge 16$, the gain saturates, and $N=32$ is almost identical to $N=16$, showing diminishing marginal returns.}
\label{fig:appendix-band-n-bar}
\end{figure}

As shown in Fig.~\ref{fig:appendix-band-n-bar}, this paper adopts $N=8$ as the default configuration. This setting offers a better balance between performance and implementation complexity and is also consistent with the threshold discretization used in the later experiments.

\subsubsection{Selection of Maximum Threshold and Weights}

For the maximum threshold $Th_{\max}$, this paper sets the upper limit to 960\,KB; the rationale is explained in Appendix~\ref{appendix:config-analysis}. The results in Fig.~\ref{fig:appendix-thmax-compare} in that appendix show that as $Th_{\max}$ increases from 240\,KB to 480\,KB and then to 960\,KB, the system's loss rate, 99th-percentile buffer occupancy, and average buffer occupancy all improve consistently. This indicates that on the current platform, a larger $Th_{\max}$ enables the probabilistic scheduling mechanism to better exploit its smoothing advantage and reduces the performance loss caused by quantization saturation. Combining the appendix analysis and experimental results, we therefore adopt $Th_{\max}=960$\,KB as the default engineering configuration.

\section{Related Work}
Existing data center load balancing work can be broadly divided into three categories. The first category includes hash-based flow-level or flowlet-level mechanisms, such as ECMP, CONGA, HULA, LetFlow, FlowBender, and PLB\cite{ECMP,CONGA,HULA2016,LetFlow2017,FlowBender2014,PLB2022}. These methods are easy to deploy, incur low control overhead, and naturally preserve flow-level ordering or amortize path changes over flowlets, but they lack sensitivity to instantaneous queue states and path-capability differences at pure packet granularity. As a result, they still tend to suffer from hotspot congestion and long-tail latency when multipath Clos networks face AI-training micro-burst traffic. The second category consists of fine-grained scheduling methods based on queue or congestion state, including micro-load balancing schemes such as DRILL\cite{Drill}, packet-spraying approaches such as DRB and Presto\cite{DRB,Presto}, and more general packet-level or transport-level designs such as pFabric, PIAS, pHost, NDP, Homa, Swift, DCQCN, IRN, ExpressPass, and DART\cite{pFabric2013,PIAS2015,pHost2015,NDP2017,Homa2018,Swift2020,DCQCN2015,IRN2018,ExpressPass2016,Dart2018}. Such methods typically exploit multipath resources more effectively when state information is fresh, but in large-scale networks they become constrained by both stale state information and hardware complexity.

The third category focuses further on robustness and engineering deployability under non-ideal network conditions, such as Hermes for asymmetric links, failures, and black-hole scenarios\cite{Hermes}, and SGLB for AI clusters with global congestion awareness\cite{SGLB}. Overall, the key challenge of modern data center load balancing is no longer merely whether congestion can be sensed, but how to realize stable, efficient, and scalable scheduling under stale state information, path heterogeneity, and limited on-chip resources. Compared with these prior efforts, PSP adopts a ``local quantization + probability mapping + weighted selection'' design, replacing global best-path search with local probabilistic dispatching. It retains state awareness while reducing implementation complexity and improving robustness to stale state information and heterogeneous-link scenarios.

\section{Conclusion}

This paper presents PSP, a packet-level load balancer for stale-state and bandwidth-asymmetric AI data center networks. PSP replaces global sorting with Band discretization, probabilistic weight mapping, and parallel weighted selection, reducing implementation complexity while retaining state awareness. Cycle-accurate simulations show that PSP mitigates scheduling concentration and oscillation under stale information, avoids bandwidth-constrained paths, and offers a better overall performance--hardware-cost trade-off than Top-$k$. These results make PSP a practical design for high-performance AI networks.

\section*{Acknowledgment}
The authors thank the anonymous reviewers for their constructive comments and suggestions.

\clearpage
\IEEEtriggeratref{18}
\bibliographystyle{IEEEtran}
\bibliography{sample}

\clearpage
\appendices
\section{Rationale for the Experimental Configuration}
The rationale behind the chosen experimental settings is as follows:
\begin{itemize}
\item \textbf{100\% offered load.} This paper targets micro-burst communication in AI training, where traffic can momentarily approach link capacity during synchronization phases. Using 100\% offered load therefore captures realistic burst pressure more accurately and more clearly differentiates the congestion-control and load-balancing capabilities of different algorithms.
\item \textbf{10\,ms simulation duration.} Micro-bursts typically occur on a millisecond timescale. A 10\,ms simulation window therefore covers an entire burst process while providing enough samples for queue-depth, loss-rate, and tail-metric statistics.
\end{itemize}

\section{Advantages and Positioning of the In-House Simulation Platform}
\label{appendix:platform}
It is important to note that our in-house platform is not intended to replace RTL-level circuit verification, nor does it directly answer timing-closure questions under specific processes, layouts, and target frequencies. Instead, this paper focuses on a layer between general-purpose network discrete-event simulation and circuit-level implementation, namely \textbf{switch-chip scheduling-pipeline modeling under hardware budget constraints}. For the packet-level load balancing problem studied here, modeling only link propagation, port queues, and control-plane state update delay is insufficient, because algorithmic performance depends not only on congestion feedback in the network, but also on whether the scheduler can complete state collection, local quantization, result aggregation, and global selection within a limited processing budget.

Unlike general-purpose network simulation approaches that abstract scheduling logic as an ``instantaneously available global decision function,'' our platform explicitly models the multi-stage internal processing of the scheduler, including the degree of parallelism in port-state quantization modules, the set of ports handled by each module, the clock interval of quantization operations, the storage and visibility delay of local quantization results, the scanning and transmission process from local modules to the global decision module, the inter-module communication latency in clock cycles, and the number of results the global decision module can process per unit time. These factors jointly determine how fresh the state observed by the scheduler is and whether the algorithm remains stable and scalable at large port counts.

Therefore, the reason for using our own platform is not to deny the value of general-purpose simulators such as ns-3, but because they answer a different level of question. ns-3 is better suited for evaluating links, queues, protocol interactions, and end-to-end network performance. In contrast, we also need to answer the following question: under a given module partition, scanning bandwidth, pipeline delay, and local aggregation constraint, can a load balancing algorithm still maintain its expected performance? If this internal processing budget is ignored and all port states are assumed to be globally visible and immediately usable at zero cost and zero delay, then algorithms that rely on large-scale global scanning or centralized selection are easily overestimated, while the value of designs based on local processing, parallel mapping, and gradual aggregation is underestimated.

Based on this consideration, our platform treats \textit{internal scheduler processing delay} as one component of stale state information and brings the chip-internal time cost, which is usually ignored by traditional network-layer simulation, into a unified evaluation framework. Although the resulting evaluation does not replace circuit synthesis or timing analysis, it more faithfully reflects the trade-off among implementation cost, information freshness, and network performance for different load balancing algorithms under finite hardware budgets. This is precisely why we choose an in-house platform rather than using a general-purpose network simulator as the sole experimental tool.

\section{Parameter Configuration Analysis}
\label{appendix:config-analysis}

\begin{figure*}[!t]
    \centering
    \subfloat[99th-percentile buffer]{%
    \begin{minipage}{0.48\textwidth}
        \centering
        \begin{tikzpicture}
            \begin{axis}[
                ybar,
                bar width=4pt,
                width=\linewidth,
                height=3.8cm,
                ylabel={99th-Percentile Buffer (MB)},
                symbolic x coords={128,256,512},
                xtick=data,
                xlabel={Number of GPUs},
                ymin=0, ymax=18,
                ymajorgrids=true,
                grid style=dashed,
                enlarge x limits=0.18,
                tick label style={font=\footnotesize},
                label style={font=\footnotesize},
                legend style={at={(0.5,1.18)}, anchor=south, legend columns=3, font=\footnotesize},
            ]
                \addplot coordinates {(128,15.95) (256,15.84) (512,15.62)};
                \addplot coordinates {(128,15.89) (256,15.59) (512,15.37)};
                \addplot[fill=blue!60] coordinates {(128,3.55) (256,3.92) (512,4.53)};
                \legend{240 KB, 480 KB, 960 KB}
            \end{axis}
        \end{tikzpicture}
    \end{minipage}}\hfill
    \subfloat[Average buffer]{%
    \begin{minipage}{0.48\textwidth}
        \centering
        \begin{tikzpicture}
            \begin{axis}[
                ybar,
                bar width=4pt,
                width=\linewidth,
                height=3.8cm,
                ylabel={Average Buffer (MB)},
                symbolic x coords={128,256,512},
                xtick=data,
                xlabel={Number of GPUs},
                ymin=0, ymax=2,
                ymajorgrids=true,
                grid style=dashed,
                enlarge x limits=0.18,
                tick label style={font=\footnotesize},
                label style={font=\footnotesize},
            ]
                \addplot coordinates {(128,1.52) (256,1.03) (512,0.93)};
                \addplot coordinates {(128,1.34) (256,0.94) (512,0.76)};
                \addplot[fill=blue!60] coordinates {(128,0.21) (256,0.19) (512,0.17)};
            \end{axis}
        \end{tikzpicture}
    \end{minipage}}
    \caption{Experimental comparison under different maximum-threshold settings $Th_{\max}$. Larger thresholds yield better performance.}
    \label{fig:appendix-thmax-compare}
\end{figure*}
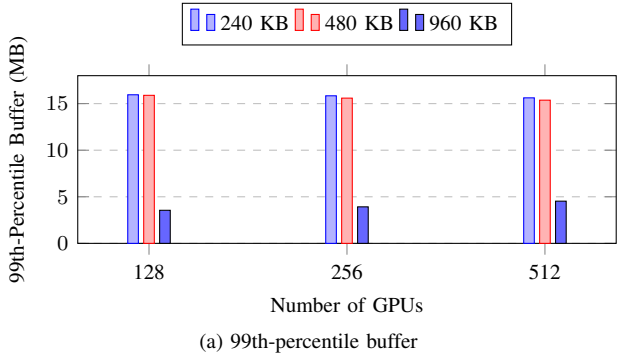
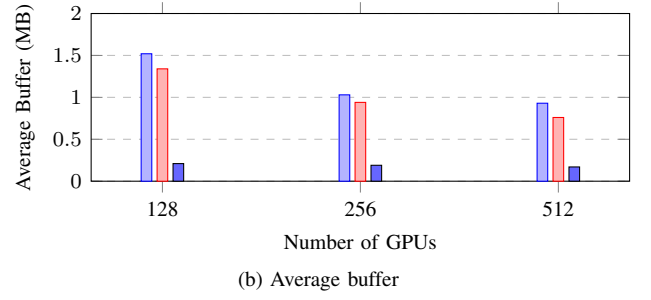

Figure~\ref{fig:appendix-thmax-compare} summarizes the measured effect of the maximum-threshold setting.

\subsection{Linear Threshold-Probability Mapping}
\textbf{The latency trap of exponential weights.}

In probability design, we first compare linear probabilities with exponential probabilities, for example by setting the probability of the $n$-th Band to $2^n / \sum 2^i$. Intuitively, exponential probabilities provide a wider control range, but analysis shows that they introduce a clear latency trap in practical hardware.

\textbf{Illustrative derivation.} Assume that all ports in the system are fully loaded (100\% throughput) and stay in Band 0, and each port has an outgoing bandwidth of 400\,Gbps. If one port drops to 50\% carrying capacity (200\,Gbps) due to a link failure or an elephant flow, then under exponential weights that congested port will quickly fall to Band 1, and its selection probability becomes 50\% of that of Band 0.

However, hardware systems inevitably incur state update delay, which we assume here to be $20\,\mu\text{s}$. When the link failure is repaired or the elephant flow disappears, the port capacity instantly returns to 400\,Gbps, but because of the $20\,\mu\text{s}$ feedback delay, the incoming traffic assigned to that port remains at 200\,Gbps. To preserve throughput during this $20\,\mu\text{s}$ interval and avoid link idling, the queue must contain enough buffered data to cover the rate gap. Link idling not only wastes physical bandwidth, but in distributed AI training, even microsecond-scale local delay can be amplified by collective synchronization semantics into cluster-wide compute underutilization \cite{Drill,TailAtScale,Aeon,Homa2018,Swift2020}. The required amount of buffered data to drain is
\begin{equation}
Q_{\text{drop}} = (400 - 200)\,\text{Gbps} \times 20\,\mu\text{s} = 0.5\,\text{MB}
\end{equation}

This means that under exponential weights, to avoid throughput degradation, the trigger threshold of Band 1 must be greater than 0.5\,MB. This leads to two severe consequences:

\begin{enumerate}
    \item \textbf{Significantly increased static queuing delay.} Once a port becomes congested and crosses the first threshold, it incurs at least a fixed queuing delay of $0.5\,\text{MB} / 400\,\text{Gbps} = 10\,\mu\text{s}$.
    \item \textbf{Substantially higher end-host reordering pressure.} Under mild imbalance, for example when one port is in Band 0 and another in Band 1, two packets sent to the same destination at the same time may experience a delay gap of $10\,\mu\text{s}$, dramatically increasing NIC-side reordering pressure\cite{DCQCN2015,IRN2018,Dart2018}.
\end{enumerate}

\textbf{Linear probability design.}
To avoid the high latency and strong reordering pressure introduced by exponential distributions, we propose a \textbf{linear probability design}. Heuristic analysis suggests that when probabilities decrease linearly and thresholds increase linearly with the same proportion, the system is more likely to achieve a smooth dynamic equilibrium.

Suppose the total incoming traffic during a time window is $a$, and the system is partitioned into $n$ Bands. Under linear probability design, the traffic assigned to each Band decreases linearly: the $n$-th Band receives 0, the $(n-1)$-th Band receives $x$, and the first Band receives $(n-1)x$. The sum across all Bands equals the total traffic:
\[
0 + x + 2x + \dots + (n-1)x = \frac{n(n-1)}{2} x = a
\]
Hence the basic traffic unit is $x = 2a / [n(n-1)]$. Therefore, the traffic share sent to the $y$-th Band can be expressed as $(n-y)x = (n-y)\cdot 2a / [n(n-1)]$.

Port buffering is essentially the integral of the mismatch between incoming and outgoing traffic. Our optimization target is the \textbf{equal-time recovery principle}: regardless of whether the source of congestion disappears, the system should maintain a stable dynamic equilibrium; once congestion is relieved, ports in different Bands should drain their extra buffers within a similar time and return to low-load states in a synchronized manner.

Let the threshold of the $y$-th Band be $Th_y$ (in Bytes), and let the net incoming traffic assigned to that Band within one time window be $(n-y)x$ (in Bytes/window). To satisfy equal-time recovery, i.e., to keep the sum of ``current queue accumulation + future draining capability'' constant across different Bands, the following must hold:
\[
Th_1 + n \cdot x = Th_2 + (n-1) \cdot x = \dots = Th_n + 1 \cdot x = \text{Constant}
\]
Each term physically represents \textbf{current queue depth (Bytes) + number of windows needed to drain at the maximum recovery rate $\times$ traffic share per window (Bytes/window)}. Since $x$ is also measured in Bytes, both sides are dimensionally consistent. After simplification, we see that the thresholds $Th$ must form an arithmetic sequence with common difference $x$. In other words, thresholds should increase linearly with Band level. The resulting configuration template is:
\begin{itemize}
\item \textbf{Threshold distribution:} $Th/n,\, 2Th/n,\, 3Th/n,\, \dots,\, n \cdot Th/n$
\item \textbf{Probability distribution:} $n/\mathit{Sum},\, (n-1)/\mathit{Sum},\, \dots,\, 1/\mathit{Sum}$, where $\mathit{Sum} = n(n+1)/2$
\end{itemize}

\subsection{Engineering Model for the Maximum Threshold $Th_{\max}$}
The maximum threshold directly determines the upper bound of network queuing delay. Different network topologies and end-host applications differ in their sensitivity to delay, while feasible thresholds are strongly constrained by the maximum reordering capability of the receiving NIC. If grid search is applied to every heterogeneous network, the tuning cost in both time and computation becomes enormous.

Based on the analysis above, we propose an engineering-oriented threshold configuration model. In heterogeneous networks, the tolerance limit of the system is mainly determined by the capacity of the receiver-side NIC reorder buffer. A reasonable range for the maximum threshold can be estimated by
\[
L_{buffer} = C_{port} \times \Delta T_{max}
\]
where $L_{buffer}$ is the maximum reorder-buffer length at the NIC, $C_{port}$ is the port throughput, and $\Delta T_{max}$ is the maximum tolerable path-delay difference.

Since the delay difference in our system is mainly determined by the queuing delay caused by maximum queue occupancy, we have $\Delta T_{max} = Th_{max} / C_{port}$. Substituting this into the above equation yields
\[
Th_{max} = L_{buffer}
\]
This shows that under the linear PSP framework proposed in this paper, the maximum threshold need not be determined by trial and error. Instead, it should be directly aligned with the hardware reordering capability of the end-host NIC, thereby minimizing tuning cost while avoiding retransmissions caused by excessive reordering. Although we do not directly measure the cycle-by-cycle occupancy of the NIC-side reorder buffer, in packet-level scheduling the dominant source of reordering pressure is the queuing-delay difference across paths, which is itself mainly determined by switch queue occupancy. For systems with fixed port rate and scheduling granularity, queue depth, queuing delay, and reordering depth are monotonically related. Therefore, this paper uses queue-buffer occupancy as an indirect metric for reordering pressure\cite{Presto,DRB,IRN2018,Dart2018}. This metric is not equivalent to actual NIC reorder-buffer occupancy, but it can stably reflect the relative impact of different scheduling algorithms on reordering risk.

It should also be noted that the derivation above implicitly assumes that single-hop queuing delay is the only source of end-to-end queuing delay. In a real two-layer Clos architecture, however, end-to-end queuing delay is the sum of queuing delay at the source leaf and at the spine. Therefore, if $Th_{\max}$ is directly set to the full capacity of the NIC reorder buffer, the accumulated reordering depth under simultaneous high occupancy at both switch layers may exceed the receiver's limit and trigger retransmissions. Considering that AI training traffic, such as RoCEv2 over DDP, is highly sensitive to reordering, engineering practice must reserve a safety margin\cite{DCQCN2015,IRN2018,ExpressPass2016,Dart2018}. That is, $Th_{\max}$ should be set to a reasonable subset of the NIC reorder-buffer capacity so that the total reordering depth after multi-hop queuing accumulation remains within the tolerance of the NIC. Based on these considerations, our experiments use $Th_{\max}=960$\,KB as a conservative yet practical default value that balances two-layer queuing accumulation and AI-traffic reordering tolerance.

\section{Illustrative Example of JSQ and PSP Under Stale Information}
\label{appendix:herding}
Consider a switch with 512 ports, each at 400\,Gbps, and a state update period of $20\,\mu\text{s}$. Assume that ports 3 and 5 are in Band 1 with weight 3, while the remaining ports are in Band 2 with weight 2. Under JSQ, because the scheduler selects only within the currently best Band, the burst traffic arriving during the $20\,\mu\text{s}$ window is highly concentrated on ports 3 and 5, causing instantaneous local congestion.

By contrast, PSP dispatches according to the global weight distribution, which smooths traffic across all candidate ports. In this way, PSP preserves preference for lower-load ports while avoiding the divergent oscillation caused by all traffic chasing a few ``optimal'' ports simultaneously.

Within this stale-information window, the total arriving traffic is
\[
512\times 400\,\mathrm{Gbps}\times 20\,\mu\mathrm{s}\times \frac{1}{8}=0.5\,\mathrm{GB}.
\]
Therefore, under PSP, the expected traffic received by ports 3 and 5 is
\[
0.5\mathrm{G}\times \frac{3}{3\times 2+2\times 510}\approx 0.00146\mathrm{G}\approx 1.46\mathrm{MB}.
\]

\section{Adaptive Regulation Mechanism Under Bandwidth Asymmetry}
\label{appendix:bandwidth-asymmetry}

Assume a system with 512 ports at $400\,\text{Gbps}$ and a state update period of $20\,\mu\text{s}$. Under 100\% offered load, suppose that the throughput capacity of one path suddenly drops from $400\,\text{Gbps}$ to $200\,\text{Gbps}$ because of a link failure.

\paragraph{Buffer buildup stage.}
During the $20\,\mu\text{s}$ state update window, the scheduler still maintains the original dispatch rate, so the rate mismatch causes buffer accumulation at the port, with an increment of
\[
\Delta Q = (400 - 200)\,\text{Gbps} \times 20\,\mu\text{s} \times \frac{1}{8} = 0.5\,\text{MB}.
\]

\paragraph{Weight suppression stage.}
This $0.5\,\text{MB}$ of backlog pushes the port state from Band 0 to a higher-load Band, for example reducing its weight from $W_{normal}=7$ to $W_{fault}=3$. The incoming traffic share assigned to that port then drops significantly. If the other normal ports keep weight 7, the relationship between the incoming traffic $x$ sent to the failed port and the traffic sent to a normal port satisfies
\[
\frac{x}{C_{normal}} \approx \frac{W_{fault}}{W_{normal}} = \frac{3}{7},
\]
from which we obtain
\[
x \approx \frac{3}{7} \times 400\,\text{Gbps} \approx 171.4\,\text{Gbps}.
\]

\paragraph{Dynamic recovery and equilibrium.}
At this point, the incoming traffic ($171.4\,\text{Gbps}$) is below the residual physical bandwidth ($200\,\text{Gbps}$), so the port queue starts to drain.
In the next sampling period, as the queue decreases, its weight may rise to $W=4$. The incoming traffic then increases to $\frac{4}{7} \times 400\,\text{Gbps} \approx 228.6\,\text{Gbps}$. Since this again exceeds the physical limit, the queue starts to build up once more.

Through this cycle, PSP uses queue depth as a ``pressure gauge'' and drives traffic into a periodically convergent oscillation near the constrained port,
 so that the average incoming rate statistically approaches the residual bandwidth limit. 
 The cost of this adaptation is a certain amount of buffer space, 
 namely the minimum backlog required to store the negative-feedback signal, together with extra queuing delay. However, 
 by setting the maximum threshold $Th_{max}$ appropriately, this cost can be kept within the reorder tolerance of the end-host NIC.

\end{document}